\documentclass[preprint]{aastex}
\usepackage{graphicx}
\usepackage{natbib}

\shorttitle{The 2008 outburst of EX\,Lup}
\shortauthors{{Juh\'asz} et al.}

\begin{document}


\title{The 2008 outburst of EX\,Lup - Silicate crystals in motion}


\author{A. Juh\'asz\altaffilmark{1,4}, 
  C.P. Dullemond\altaffilmark{1},  
  R. van Boekel\altaffilmark{1}, 
  J. Bouwman\altaffilmark{1}, 
  P. \'Abrah\'am\altaffilmark{2}, 
  J.A. Acosta-Pulido\altaffilmark{3},
  Th. Henning\altaffilmark{1},
  A. K\'osp\'al\altaffilmark{4}, 
  A. Sicilia-Aguilar\altaffilmark{1}, 
  A. Jones\altaffilmark{5},
  A. Mo\'or\altaffilmark{2}, 
  L. Mosoni\altaffilmark{2}, 
  Zs. Reg\'aly\altaffilmark{2}, 
  Gy. Szokoly\altaffilmark{6},
  N. Sipos\altaffilmark{2}
  } 
  \email{juhasz@mpia-hd.mpg.de}
\altaffiltext{1}{Max-Planck-Institut f\"ur Astronomie, K\"onigstuhl 17, Heidelberg, D-69117 Germany}
\altaffiltext{2}{Konkoly Observatory of the Hungarian Academy of Sciences, PO Box 67, H-1525, Budapest, Hungary}
\altaffiltext{3}{Instituto de Astrofis\'{i}ca de Canarias, E-38200 La Laguna, Tenerife, Canary Islands, Spain}
\altaffiltext{4}{Leiden Observatory, Leiden University, P.O. Box 9513, 2300 RA Leiden, The Netherlands}
\altaffiltext{5}{31 Ranui Road, Stoke, Nelson 7011, New Zealand}
\altaffiltext{6}{E\"otv\"os Lor\'and University, Inst. of Physics, P\'azm\'any P. s. 1/A, H-1117 Budapest, Hungary}



\begin{abstract}

EX\,Lup is the prototype of the EXor class of eruptive young stars. These objects show optical outbursts which are 
thought to be related to runaway accretion onto the star. In a previous study we observed in-situ crystal formation in 
the disk of EX\,Lup during its latest outburst in 2008, making the object an ideal laboratory to investigate circumstellar 
crystal formation and transport. This outburst was monitored  by a campaign of ground-based and Spitzer Space Telescope observations.
Here we modeled the spectral energy distribution of EX\,Lup in the outburst from optical to millimeter wavelengths with a 2D radiative 
transfer code. Our results showed that the shape of the SED at optical wavelengths was more consistent with a single temperature blackbody 
than a temperature distribution. We also found that this single temperature component emitted 80--100\,\% of the total accretion luminosity. 
We concluded that a thermal instability, the most widely accepted model of EXor outbursts, was likely not the triggering mechanism 
of the 2008 outburst of EX\,Lup. 
Our mid-infrared Spitzer spectra revealed that the strength of all crystalline bands between 8 and 30\,{\micron} increased right after the 
end of the outburst. Six months later, however, the crystallinity in the 10\,{\micron} silicate feature complex decreased. 
Our modeling of the mid-infrared spectral evolution of EX\,Lup showed that, although vertical mixing should be stronger during the outburst than 
in the quiescent phase, fast radial transport of crystals (e.g., by stellar/disk wind) was required to reproduce the observed mid-infrared spectra.

\end{abstract}



\section{Introduction}
\label{sec:introduction}

The EXor class of young eruptive stars, named after the prototype EX\,Lup, was established on the
basis of the peculiar optical behaviour of these sources \citep{ref:herbig1977}. EXors are known to show 
repetitive optical brightenings separated by faint quiescent periods. 
Typical time-scales of the bright states range from several months to few years. The
time between two subsequent outbursts is typically several years. During the outbursts the brightness
of the source increases by 2--5\,mag in the V-band. In the quiescent phase the optical 
spectra of these stars resemble that of typical low-mass T\,Tauri stars with a spectral type 
of late K to early M. In the bright state, however, the optical spectra are dominated by broad emission lines usually
seen in very early-type stars \citep{ref:herbig2008}. The excess emission above the stellar photosphere at infrared 
wavelengths is also a characteristics of the class which is attributed to a protoplanetary accretion disk 
\citep{ref:hartmann_kenyon1996}. 

In general, optical outbursts of EXors are thought to be related to runaway accretion from the disk 
onto the young star \citep{ref:hartmann_kenyon1996}. This scenario is supported by observations of accretion tracers (e.g. spectral 
lines, veiling) which indicate higher mass accretion rate in the high phase compared to the quiescent 
periods (e.g. \citealt{ref:lorenzetti2007}). Little is known, however, on the details of the mechanism(s) leading to the increase 
of the accretion rate. It is generally believed that EXor outbursts are caused by the same kind of disk
instabilities as those driving the eruptions in FU\,Ori type objects (EXors are sometimes 
also called sub-FUors) \citep{ref:herbig2008}. 

FUor and EXor outbursts can also play an important role in the thermal processing of protoplanetary dust.
During outbursts, the enhanced accretion and irradiation luminosity in the central regions of the system should 
lead to a strong increase of the dust temperature in the disk. Wherever the temperature exceeds a threshold value, the 
initially amorphous silicate grains crystallize rapidly. Crystalline silicates are not present in the interstellar 
medium \citep{ref:kemper2005}, from where the dust content of protoplanetary disks originates. Sharp features of silicate crystals at mid-infrared 
wavelengths are, however, frequently observed towards young stars with various spectral types (e.g., \citet{ref:van_boekel2005, ref:bouwman2008, 
ref:watson2009, ref:meeus2009} or see, Henning \& Meeus (2010) for a review), as well as solar system comets 
(e.g., \citet{ref:wooden2007} and references therein). 
Although these observations provide evidence for crystal formation in protoplanetary disks, the process itself was not actually
observed until \citet{ref:abraham2009} reported the first observation of in-situ crystal formation in its recent outburst of 
EX\,Lup.

Crystalline silicates are also observed in the cold (T$<$300\,K) outer disk, where the temperature is too low
for crystallization (e.g., \citealt{ref:bouwman2008, ref:watson2009, ref:meeus2009}). Cometary bodies, which can contain high abundance 
of crystalline silicates,  are also formed in this cold part of protoplanetary disks. 
An important question is whether crystals were formed in situ in the outer disk (e.g. \citealt{ref:desch2005}) or were formed in the 
central region of the system at high temperature and subsequently transported outwards (e.g. \citealt{ref:gail2004}). 
In the lack of time-resolved observations it is hard to distinguish between the two scenarios. In the case of EX\,Lup, however, 
\emph{we know} that the crystals are formed in the hot (T$>1000$\,K) inner regions of the system \citep{ref:abraham2009}. Thus, the 
recent outburst of EX\,Lup provides the unique opportunity to use the silicate crystals to trace the transport mechanisms in the disk of EX\,Lup. 

EX\,Lup is a young (3\,Myr) low-mass (0.6\,M$_\odot$) M0 star \citep{ref:gras_velazquez_ray2005}
located in the Lupus cloud complex. In the 1950s, EX\,Lup had a large outburst ($\Delta V\approx$5\,mag), but 
only smaller amplitude brightenings, separated by few years of quiescent periods, were observed ever since. EX\,Lup has an 
infrared excess above the stellar photosphere longwards of about 2\,{\micron}, which originates in a protoplanetary disk. 
So far no evidence for an envelope has been found and indeed \citet{ref:sipos2009} successfully modeled the quiescent SED of 
EX\,Lup using a protoplanetary disk model only. Apart from a large inner hole in the dust disk, no other observational difference has been 
found which would distinguish EX\,Lup from other T\,Tauri stars. The 2008 outburst of EX\,Lup was observed with various ground-based 
instruments as well as with the Spitzer Space Telescope. In this paper we analyze these data in order to constrain
dust processing and the outburst mechanism.


\section{Observations}
\label{sec:observations}

In our observation campaign we used a variety of ground- and space-based telescopes/instruments. We attempted 
to collect as co-temporary measurements as possible to build the SED of EX\,Lup in the outburst. In the following
we will describe the details of the data reduction for each used instrument, while the log of our observations
as well as the resulting flux densities are listed in Tab.\,\ref{tab:logobs}.

\subsection{Optical light curve} 
\begin{figure}[!ht]
\includegraphics[scale=0.6, angle=0]{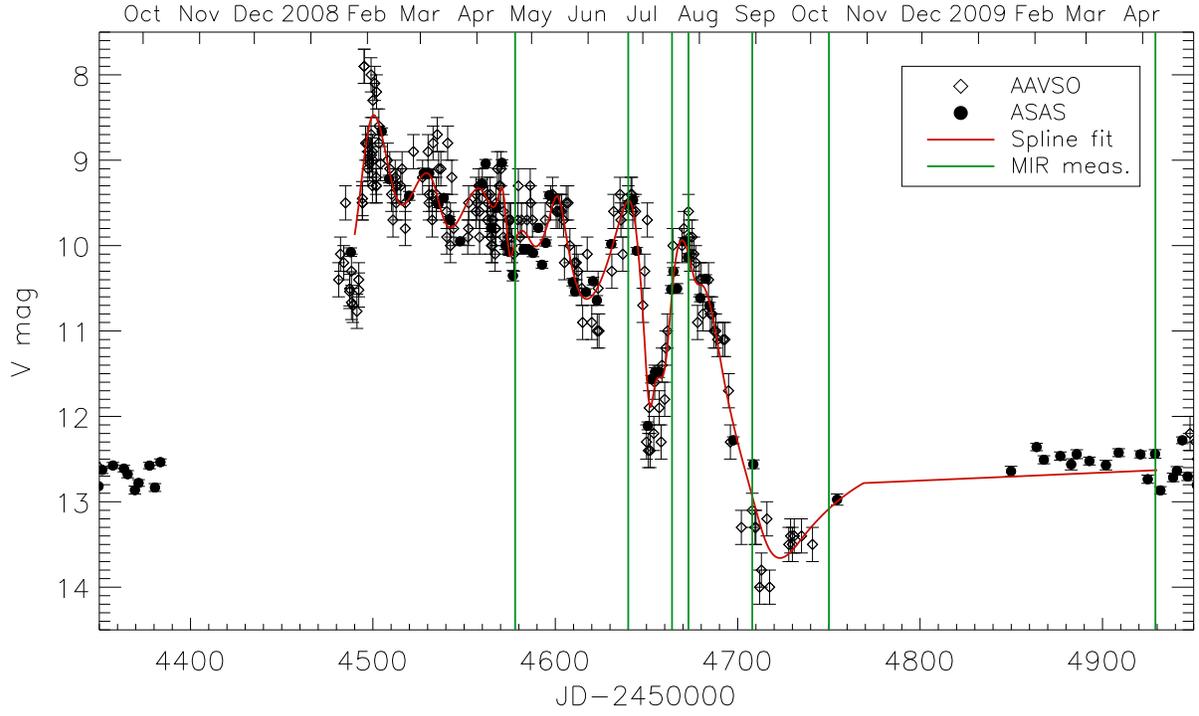}
\caption{Optical light curve of EX\,Lup. The red line in the middle of the points shows the spline fit
used for our 2D RT simulation. The green vertical lines mark the epochs when mid-infrared spectra were
taken. }
\label{fig:light_curve_pure}
\end{figure}

The optical brightness variations during the course of the outburst were monitored
by amateur and professional astronomers. We collected visual brightness estimations from Albert Jones, the 
AAVSO\footnote{www.aavso.org} database and V-band CCD observations from the ASAS-3 database \citep{ref:pojmanski2002}. 
For the visual magnitude estimation we used an uncertainty of 0.3\,mag, which was the average difference between
EX\,Lup and the comparison stars. From the ASAS-3 database we used only the highest quality measurements (quality flag 'A'), 
for which the uncertainty of the measured V-band magnitude was less than 0.1\,mag. The optical light curve is
bpresented in Fig.\,\ref{fig:light_curve_pure}. 

\subsection{GROND}
We observed EX\,Lup with the Gamma-Ray Burst Optical and Near Infrared Detector (GROND), which is a 7-channel
simultaneous optical-infrared imager at the ESO 2.2\,m telescope, on 20 April 2008. Optical images, covering a field of 
view of ${5.4'}\times {5.4'}$, were obtained in the $g',r',i',z'$ bands. The total observing sequenced consisted of 4
dithered exposures with 1\,s integration time. At near-infrared wavelengths (JHK$_S$) images were obtained in 4 (JH) or 
24 (K$_S$) dither positions, with a total of 40 sec exposure time. The field-of view was ${10'}\times {10'}$.
The optical images were reduced with standard IRAF tasks, and the final magnitudes were determined with aperture 
photometry and adopting the nominal zeropoint of each optical band from \citep{ref:greiner2008}. Near-IR magnitudes 
were determined by averaging the photometric zeropoints derived for several stars in the field using the 2MASS 
catalog. Because the near-infrared images were defocused, a large aperture was adopted in order to include all flux.
The estimated photometric accuracy is 2.5, 2.5 and 5\% in the J, H, and K$_S$ band, respectively, while in the optical
the uncertainty is significantly larger (15\,\%) due to the lack of dedicated calibration measurements.

\subsection{WFI}
EX\,Lup was observed with the Wide Field Imager (WFI) on the 2.2m telescope in
La Silla on 20 April 2008, with about one hour difference with respect
to the GROND observations. The field of view of WFI spans 33$'\times$34$'$
with small holes between the 8 CCD detectors. EX\,Lup
was observed using standard Johnson UBV filters. A total of 3 exposures
were taken with each filter, with individual exposure times of 4\,s for U and 0.5\,s 
for BV. The data was reduced using standard IRAF procedures.

The  photometry was done as standard aperture photometry with a 15 pixel
aperture. The calibration and aperture corrections were obtained from
observations of the Landolt fields SA107 and SA110 (Landolt 1992) observed
at 6 different airmasses, resulting in zeropoint errors $<$0.01\,mag.  

Optical photometries corrected for emission line contamination are also listed in Tab.\,\ref{tab:logobs}. 
Correction was done by convolving optical spectra (Sicilia-Aguilar et al., in prep)
with the WFI and the GROND filter curves and calculating the ratio of the total observed emission
over the continuum. The emission line  corrected photometries were used during the
SED fitting, since our model did not include optical emission lines. 

\subsection{NTT SOFI}
EX Lup was observed on 20 April 2008 using SOFI at the ESO 3.5\,m New
Technology Telescope (NTT).  We obtained near infrared images using the narrow band 
filters NB1.215, NB1.71 and NB2.195. The observations were performed using the 
template {\it img\_obs\_AutoJitter}, taking five offsets, and repeating the 
exposure several times at each jitter position. The high brightness of the target prevents
to use commom broad band filters. Individual frames were taken using the
minimum possible exposure time of 1.2s. The total exposure time per filter was 30s in
all cases. The data were reduced using the SOFI pipeline provided by ESO
together with the Gasgano file manager. Photometric calibration was obtained by 
comparison with field of view stars included in the 2MASS catalog. The zero points 
were derived using between 6 to 8 stars with accurate 2MASS magnitudes, the 
standard deviation was about 0.03 mag in all cases.        

In addition, SOFI was used to obtain a spectrum using the high resolution 
grism covering the K band with a slit width of 1". A total exposure time 
of 560~s was dedicated to the EX Lup spectrum, split into 8 individual
frames of 70s each. A close solar type star (HIP78466) was observed to correct 
for telluric absorption. The data were also reduced using the SOFI pipeline. 

The SOFI and the GROND near-infrared photometries agree with each other
within the uncertainties in the H and K bands. In the J band the difference is about
3.8\,$\sigma$, which indicate real difference, however with low significance. 
Such difference can be caused by short time-scale (several hours) 
low amplitude intrinsic photometric variability. 

 The total spectral line contamination in EXors is usually weaker at near-infrared 
wavelengths compared to the optical (see also e.g. \citealt{ref:lorenzetti2007}). To verify
this in the case of EX\,Lup we used the SOFI K-band spectrum for the NB2.195 filter. Since
for the remaining two filters we did not have simultaneous spectroscopic data we used the 
spectra from \citep{ref:kospal2011} for the NB1.71 and NB1.215 filters, respectively. 
Although the data in \citep{ref:kospal2011} were taken three months later than our photometry, 
the optical brightness of the source was about the same at both epochs and we assumed that the
strength of the line emission was also similar on these dates. 
Since the observed line contamination was less than 1\% in all three filters, which is significantly
less than the photometric accuracy, we did not correct the photometries for line contamination.

\subsection{SPITZER IRS}
EX\,Lup was observed with the Infrared Spectrograph (IRS; \citealt{ref:houck2004}) 
onboard the Spitzer Space Telescope at four epochs; 18 March 2005 (PID: 3716), 
21 April 2008 (PID: 477), 10 October 2008 (PID: 524) and 7 April 2009 (PID: 524). 
The data of the first two epochs were published in \citet{ref:abraham2009} and 
we refer to that paper for the details of the data reduction procedure of those
spectra. 

At the last two epochs EX\,Lup was measured using only the low-resolution 
modules (R=60-120). Our spectra are based on the {\tt droopres} products processed 
through the S15.3.0 version of the Spitzer data pipeline. 
First, the background has been subtracted using associated pairs of imaged spectra from the two 
nod positions, also eliminating stray light contamination and anomalous dark currents. 
Pixels flagged by the data pipeline as being "bad" were replaced with a value interpolated 
from an 8 pixel perimeter surrounding the flagged pixel. The spectra were extracted using 
a 6.0 pixel and 5.0 pixel fixed-width aperture in the spatial dimension for the Short Low 
and the Long Low modules, respectively. The low-level fringing at wavelengths $>$20\,{\micron}
was removed using the irsfringe package \citep{ref:lahuis_boogert2003}. The spectra were 
calibrated with a spectral response function derived from IRS spectra and MARCS stellar 
models for a suite of calibrators provided by the Spitzer Science Centre. To remove any 
effect of pointing offsets, we matched orders based on the point spread function of the IRS 
instrument, correcting for possible flux losses. 

\subsection{SPITZER MIPS}
We obtained photometry at 70\,{\micron} and a low resolution spectrum of EX\,Lup,
using the Multiband Imaging Photometer for Spitzer (MIPS; \citealt{ref:rieke_2004}). 
The 70\,{\micron} imaging was performed in photometry mode. 
The data reduction was started with the Basic Calibrated Data (BCD) products 
generated by the pipeline (version S17.2) developed at the {\sl Spitzer Science Center}. 
We performed column spatial filtering and time median filtering on the BCD images 
following \citet{ref:gordon_2007}. 
The filtered BCD data were coadded and corrected for array distortions with the
SSC MOPEX (MOsaicking and Point source Extraction, \citealt{ref:makovoz_marleau2005})
software. Outlier pixels were rejected using a 3\,$\sigma$  clipping threshold.
The output mosaic was resampled to 4\arcsec pixel$^{-1}$.

We used a modified version of the IDLPHOT routines to detect the source on the final image
and to obtain aperture photometry with an aperture radius of 16{\arcsec} and with  
sky annulus between 18{\arcsec} and 39{\arcsec}. 
The aperture correction, appropriate for a 60\,K blackbody, was taken from \citet{ref:gordon_2007}.
The final uncertainties were computed by quadratically adding the internal error and the absolute calibration 
uncertainty for which we adopted 7\% (see MIPS Data Handbook). 

The low-resolution far-IR (55--95$\mu$m; R$\sim$ 15--25 ) spectrum of EX\,Lup was 
obtained in the spectral energy distribution (SED) mode of MIPS. Six observing cycles were performed, each 
consisting of 10\,s long on- and off-source exposures. The on-source and the background positions were separated by 1\arcsec. 
We began our data reduction with the BCD images (pipeline version S17.2) and MOPEX was 
used to perform further processing steps (combination of data, background removal, application of the dispersion solution).
The spectrum was extracted in an 5\,pixel wide aperture. The aperture correction factors were taken from \citet{ref:lu2008}.

\subsection{VLTI MIDI}

EX\,Lup was observed with MIDI \citep{ref:leinert2003}, the N-band instrument on the VLTI interferometer 
\citep{ref:glindemann2000}. Spectrally resolved 
interferometric visibilities were obtained between 8 and 13~$\mu$m with a spectral resolution of $R$\,$\approx$\,35. 
Observations were performed at two epochs, during the nights starting 21 June 2008 (baseline U2U3, projected length 
37.4\,m, position angle 57.6$^\circ$ E of N) and 16 July 2008 (baseline U1U4, projected length 121.1\,m, position angle 
73.4$^\circ$ E of N). Data were obtained in "high sense" mode, i.e. each interferometric observation in which the 
beams coming from both telescopes are combined coherently is followed by photometric observations in which first only 
the light coming from telescope A, then from telescope B, is measured, and the intensities of the individual beams are determined.

Calibration stars were observed to determine the system response and atmospheric transparency. During the night of 
21 June 2008 the calibrator HD~149447 (K6III, $\theta$=4.68$\pm$0.05\,mas, \citealt{ref:cohen1999}) was observed 
immediately following the EX\,Lup observation. During the night of 16 July 2008 the same calibrator was observed 
immediately before and after the EX\,Lup observations. We used the EWS package to reduce the data, and verified that 
reductions using MIA yield consistent results.

\subsection{VLT VISIR}

Low-resolution (R$\approx$250) N-band spectroscopy of EX\,Lup was performed with VISIR at the VLT. We observed the 
source during two epochs, first around 25 July 2008 (for which data from the nights of the 24, 25, 
and 27 of July were combined), and second in the night of 28 August 2008. Standard chopping and nodding 
techniques were applied to correct for the high celestial and instrumental background inherent to mid-IR observations. 

During the July observations, HD~149447 was observed as telluric and flux calibrator immediately following each observation 
of EX\,Lup, at essentially identical airmass. During the observations in August we 
used HD~196171. Both calibrators are taken from the catalogue of \citet{ref:cohen1999}, that provides spectral types 
and angular diameters for each star, which we used to compute Kurucz model spectra to calibrate the absolute flux levels.

\subsection{APEX LABOCA}
EX\,Lup was observed at 870{\micron} using the 295-element LABOCA (Large Apex Bolometer
Camera, Siringo et al. 2009) on the 12-m APEX radio telescope (Atacama Pathfinder
Experiment, \citealt{ref:gusten2006}). The observation was performed on 21th of April
2008 in continuous integration mode using a spiral pattern. This observing mode provides 
a fully sampled map of the total field-of-view of LABOCA. 
The total on-source observing time was 2.2\,h. The data processing was performed by the 
Bolometer array Analysis Software (BoA) package following the data reduction steps (correction 
for the atmospheric opacity, flux calibration, 
removal of correlated noise using an iterative method, removal of spikes) 
outlined in the BoA User Manual (version 3.1). For the flux calibration we used the
sources IRAS16293, B13134, G34.3.The co-added final map had pixels with sizes of 6{\arcsec}.
Aperture photometry was used to extract flux from the map using the position of EX\,Lup.

\section{Results}
\label{sec:results}


\subsection{Light curve}
The brightest ever recorded outburst of EX\,Lup was announced in February 2008 \citep{ref:jones2008}. 
The light curve, composed from visual estimations and V-band observations,
is presented in Fig.\,\ref{fig:light_curve_pure}. Unfortunately the beginning of the outburst was not observed, one
can only extrapolate for the starting date. The last optical measurement (ASAS) in the quiescent phase was on 
10th Oct 2007 ($\sim$110 days before the peak brightness), when brightness was V=12.7 mag.
We estimated the total length of the outburst to be about 300 days, which is an average value within the class. 

The brightness of EX\,Lup at the peak of the outburst was about V=8\,mag. Taking the quiescent brightness
to be V$\approx13$\,mag \citep{ref:sipos2009} one gets a V-band amplitude of 5\,mag for the outburst. With this
value, the 2008 outburst of EX\,Lup belongs to the strongest ones observed in EXors. This is
an unusually high value compared also to the historical record of EX\,Lup. The only comparable outburst was observed
in the 1950s. 

The shape of the light curve during the 2008 outburst of EX\,Lup showed some peculiar features. The observations
began about three weeks before the peak of the outburst. The brightness of the source rose steeply until it
achieved a peak brightness of V=8\,mag. The star remained this bright only for a few days, then it faded to about 9.5 mag. 
Then the average brightness decreased by another 2\,mag in the next 200 days. One of the most interesting properties of 
the light curve was the quasi-periodic brightness variation seen on top of the plateau-like slow fading. The amplitude of these
fluctuations increased with time, while their period remained roughly constant, about 35 days. 


\subsection{Spectral Energy Distribution}

\begin{figure}[!ht]
\begin{center}
\includegraphics[scale=0.4, angle=0]{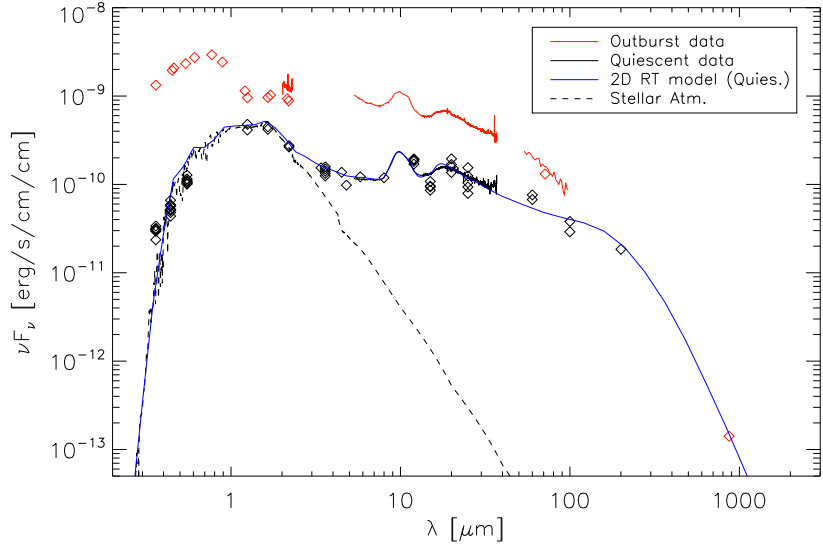}
\caption{SED of EX\,Lup in the 2008 outburst. Data shown with red are measured within 24 hours, 21 Apr 2008. 
The blue line shows the quiescent phase disk model, while the black dashed line shows the 
emission of the M0 (T$_{\rm eff}$=3800\,K) star in the quiescent phase.}
\label{fig:SED}
\end{center}
\end{figure}

The SEDs of EX\,Lup on 21 April 2008 and in the quiescent phase are compared in Fig.\,\ref{fig:SED}.  
The outburst-to-quiescent flux ratio was the highest in the U-band and it decreased gradually with wavelength.
Although EX\,Lup was already fainter on 21 Apr 2008 than at the peak of the outburst, at optical wavelengths
it was still more than an order of magnitude brighter in the outburst compared to the quiescent phase. 
The shape of the SED in the optical domain is consistent with blackbody emission at $\sim6500$\,K, which is far higher 
than the effective temperature of the star in quiescent phase (3800\,K, \citealt{ref:gras_velazquez_ray2005}). 
Although the absolute flux levels in the  GROND i and z filters are somewhat higher than what would be expected for the 
aforementioned blackbody emission, the i-z color fits well to the modeled emission. The discrepancy in the absolute flux 
level can be explained by the relatively high calibration uncertainty in the GROND optical bands. 
In the near-infrared domain the flux increased only by a factor of about 2.5 in the outburst compared to the quiescent phase.
Moving to longer wavelengths the ratio of the outburst-to-quiescent fluxes increased again to about an order of magnitude at 
5.5\,{\micron}, which is the shortest wavelength of the Spitzer IRS spectra.  Longwards of 5.5\,{\micron} the ratio of 
outburst-to-quiescent fluxes decreased with wavelength. 

\begin{figure}[!ht]
\begin{center}
\includegraphics[scale=0.4, angle=0]{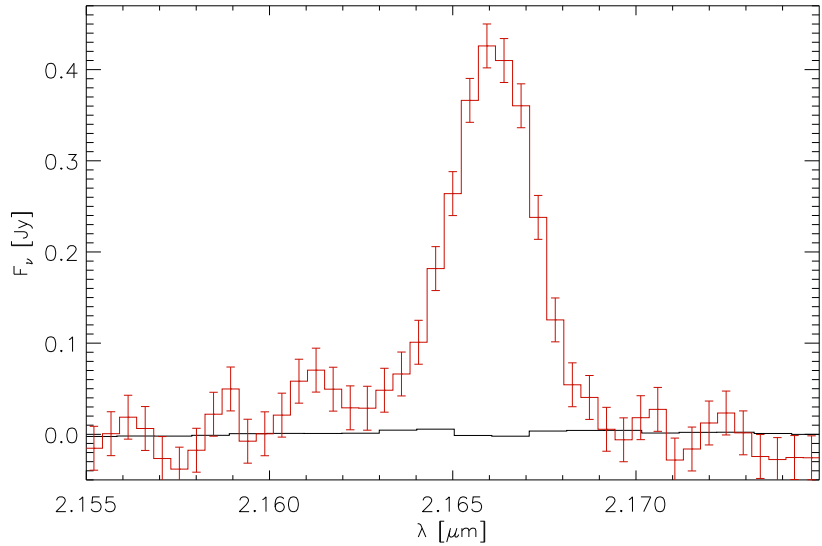}
\caption{Continuum-subtracted near-infrared spectrum of EX\,Lup around the Br\,$\gamma$ line 
in the quiescent (black solid line) and in the outburst phase (red solid line). The luminosity of
the Br\,$\gamma$ line strongly increased in the outburst compared to the quiescent phase indicating
the increase of the accretion rate. }
\label{fig:bracket_gamma}
\end{center}
\end{figure}

The brightening was certainly caused by the increased accretion rate onto the star during the outburst, indicated
by the increasing luminosity in the accretion indicator lines. In Fig.\,\ref{fig:bracket_gamma} we compare the spectral 
region around the Br\,$\gamma$ line in the outburst and in the quiescent phase from \citet{ref:sipos2009}. It can be seen
that the luminosity of the Br\,$\gamma$ was far higher in the outburst than in the quiescent phase.  
Using the relationship of \citet{ref:muzerolle1998} we derived an accretion rate of $2\cdot 10^{-7}$M$_\odot$/yr in
the outburst\footnote Note, that the hydrogen recombination lines are not only indicators for accretion, but
also mass loss. Therefore, the derived accretion rates should be considered as upper limit.
This is about three orders of magnitude higher than the value in the quiescent phase found by \citet{ref:sipos2009}.
We want to note, that different lines were used to derive the accretion rate in the outburst (Br\,$\gamma$) and 
in the quiescent (Pa\,$\beta$), since none of the accretion indicator lines were measured both in the quiescent and in 
the outburst. The usage of different accretion indicators might introduce some additional uncertainty when a comparison 
of the accretion in the outburst and in the quiescent is done. This uncertainty, however, does not affect our conclusion
that the 2008 brightening of EX\,Lup was certainly caused by the increased accretion rate onto the star during the outburst, 
indicated by the increasing luminosity in the accretion indicator lines.


\subsection{Mid-infrared spectra}

\begin{figure}[!ht]
\begin{center}
\includegraphics[scale=0.6]{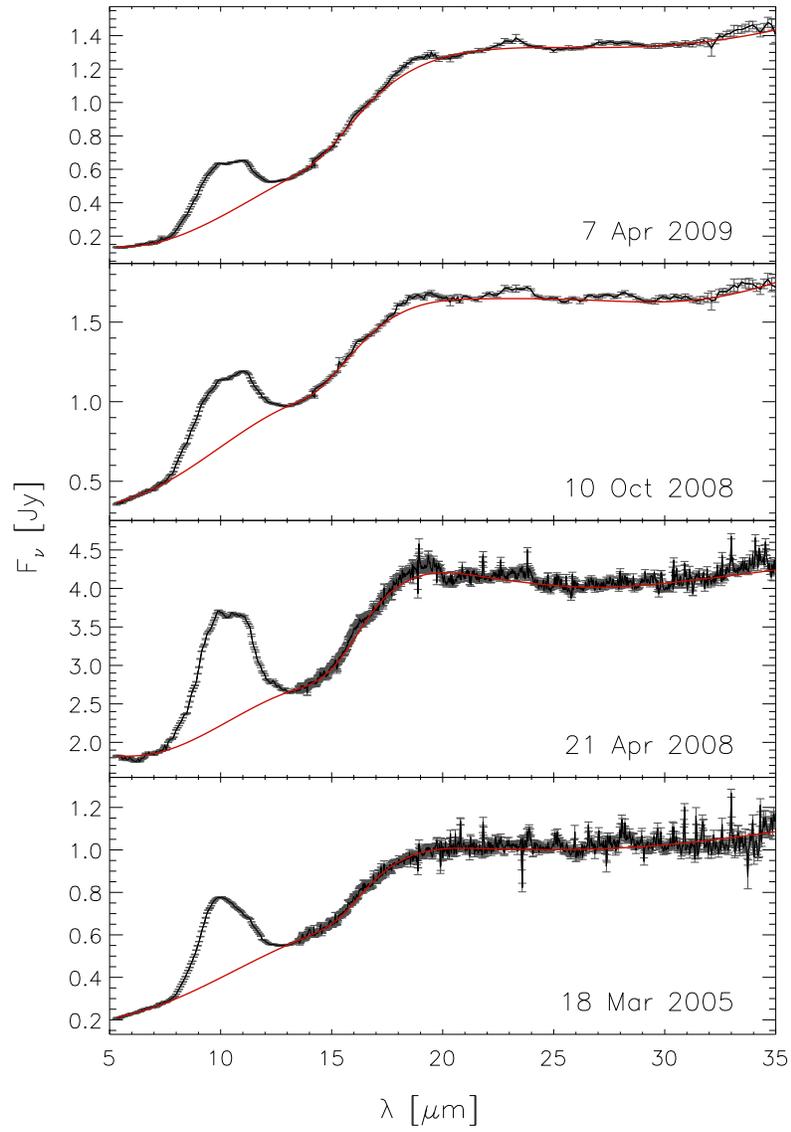}
\caption{Spitzer IRS spectra of EX\,Lup. The red line shows the fitted  continuum emission.}
\label{fig:irs_spectra_full}
\end{center}
\end{figure}

\begin{figure}[!ht]
\begin{center}
\includegraphics[scale=0.4, angle=0]{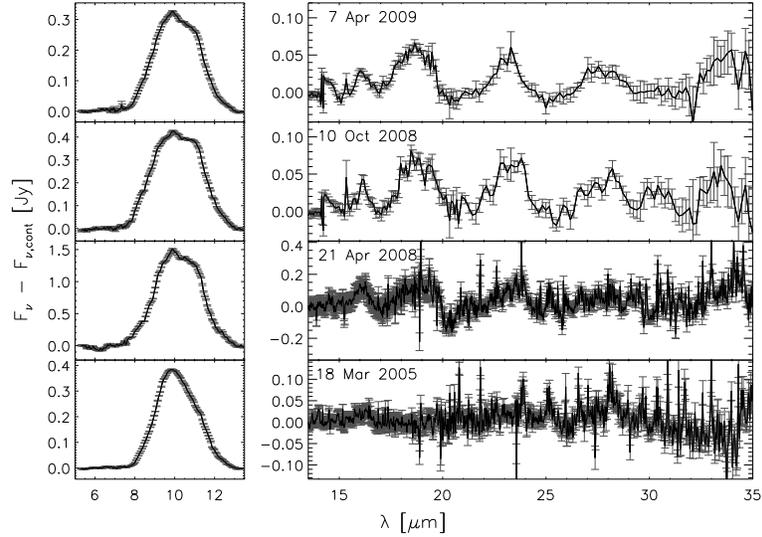}
\caption{Continuum subtracted Spitzer IRS spectra of EX\,Lup. The 5.5--13.5\,{micron} ({\it Left}) and 
13.5--35\,{\micron} ({\it Right}) wavelength intervals are shown separately with different scales in order to
show also the weaker silicate band longwards of 18\,{\micron}. }
\label{fig:irs_spectra_csub_split}
\end{center}
\end{figure}

In this section we will describe the mid-infrared spectra measured towards EX\,Lup in our observing
campaign. In order to avoid confusion, we will use the following naming convention. The ensemble
of features between 8 and 13\,{\micron} will be called the "10\,{\micron} feature complex". 
The "10\,{\micron} feature" term will be used for the forsterite band peaking at 9.9\,{\micron}. 
The Spitzer IRS spectra of EX\,Lup are presented in Fig.\,\ref{fig:irs_spectra_full}. In order to
show also the weaker dust features, a spline continuum was fitted to and then subtracted from each spectrum. 
The continuum subtracted spectra are shown in Fig.\,\ref{fig:irs_spectra_csub_split}.

The pre-outburst spectrum from March 2005 was already analyzed by \citet{ref:sipos2009}. They showed that
the dust in the disk atmosphere of EX\,Lup is dominated by small amorphous grains with a mass-weighted average
grain size of 0.5\,{\micron}. They also showed that the abundance of crystalline
silicates in the disk atmosphere is negligible. 

The Spitzer IRS spectrum from the outburst, taken on 21 April 2008, showed remarkable changes.
First of all, the absolute flux level increased by about an order of magnitude at the shortest wavelength
of the spectra (5.5\,{\micron}) and by about a factor of five at the longest wavelength ($\sim$35\,{\micron}).
The amount of brightening decreased with wavelength shortwards of about 20\,{\micron}, but it was completely 
wavelength independent longwards of 20\,{\micron}. As reported by \citet{ref:abraham2009} the spectrum 
showed a small, but narrow and sharp feature at 10\,{\micron} on top of the 10\,{\micron} feature complex 
(see Fig.\,\ref{fig:irs_spectra_csub_split}). A shoulder at 11.2\,{\micron} and another feature
at 16\,{\micron} also appeared in the outburst spectrum which were not present in the pre-outburst spectrum.
\citet{ref:abraham2009} identified the source of these new bands as crystalline forsterite. It was also 
reported that no crystalline feature was seen in the outburst spectrum longwards of 20\,{\micron}.

The third Spitzer IRS spectrum, taken on 10 October after the end of the outburst, showed further changes.
As can be seen in Fig.\,\ref{fig:irs_spectra_full}, the absolute flux level decreased compared to the
outburst spectrum. The flux level was still higher than in the quiescent spectrum by about a factor of 1.7, 
but this factor was wavelength independent in the whole Spitzer IRS wavelength domain. 
The shape of the spectral features between April 2008 and October 2008 also changed. 
In Fig.\,\ref{fig:irs_spectra_csub_split} it can be seen that the 10\,{\,micron} feature complex became 
broader and the 11.2\,{\micron} shoulder became stronger, relative to the underlying broad amorphous feature, 
in October compared to the April spectrum. The most remarkable changes occured, however, at longer wavelengths. 
Apart from the narrow 16\,{\micron} band of forsterite, new bands appeared in the October 2008 spectrum
at approximately 23\,{\micron}, 28\,{\micron} and 33\,{\micron}. These wavelengths correspond to the peak 
wavelengths of the strongest forsterite bands in this wavelength interval (see e.g. \citealt{ref:koike2003} 
for a list of forsterite band positions). Thus we identified the source of these bands to be crystalline forsterite.

The last Spitzer IRS spectrum was taken on 7 April 2009. The absolute flux level decreased further compared to the 
first post-outburst spectrum from about five months earlier. Longwards of 20\,{\micron} the flux decrease was independent 
of the the wavelength. The flux variation showed, however, clear wavelength dependence shortwards of approximately 20\,{\micron}, 
where the brightness decreased the most. The flux at 5.5\,{\micron} decreased to about half of the pre-outburst value, while
longwards of 20\,{\micron} the flux in the last IRS spectrum was approximately 30\,\% above the pre-outburst value. 
The shape of the spectral features longwards of 13\,{\micron} remained unchanged within the uncertainties
(see Fig.\,\ref{fig:irs_spectra_csub_split}), while the 10\,{\micron} silicate feature complex clearly changed compared to the previous epoch. 
It can be seen in Fig.\,\ref{fig:irs_spectra_csub_split}, that the 10\,$\mu$m forsterite feature on top of the broad amorphous 
feature was still present, but the shoulder at 11.2\,{\micron} became much weaker in April 2009 than it was in October 2008. 
The general shape of the 10\,{\micron} feature seems to be somewhere between the outburst and pre-outburst spectra. 

Due to the difference in the noise levels between the outburst and the post-outburst spectra, we cannot exclude the possibility from the 
Spitzer IRS spectra alone that weak crystalline features (of the same strength as in the post outburst spectra) longwards of 20um were already 
present in the outburst spectrum. In the post-outburst spectra the peak of the 24 and 28 um features is 9 and 6 times the corresponding flux uncertainty 
above the underlying continuum, respectively. The absolute noise level of the outburst spectrum is, however, worse than the post outburst spectra by a 
factor of about 3.4. If we measured the post-outburst spectra with the same noise level as we had in the outburst spectrum, the detection of the 
24 and 28 um features would be about 2.6 and 1.8 times the corresponding flux uncertainty, respectively, i.e. they could be hidden by
the noise. We will further discuss this question in Sec.\,\ref{sec:mixing}

\begin{figure}[!ht]
\begin{center}
\includegraphics[scale=0.4, angle=0]{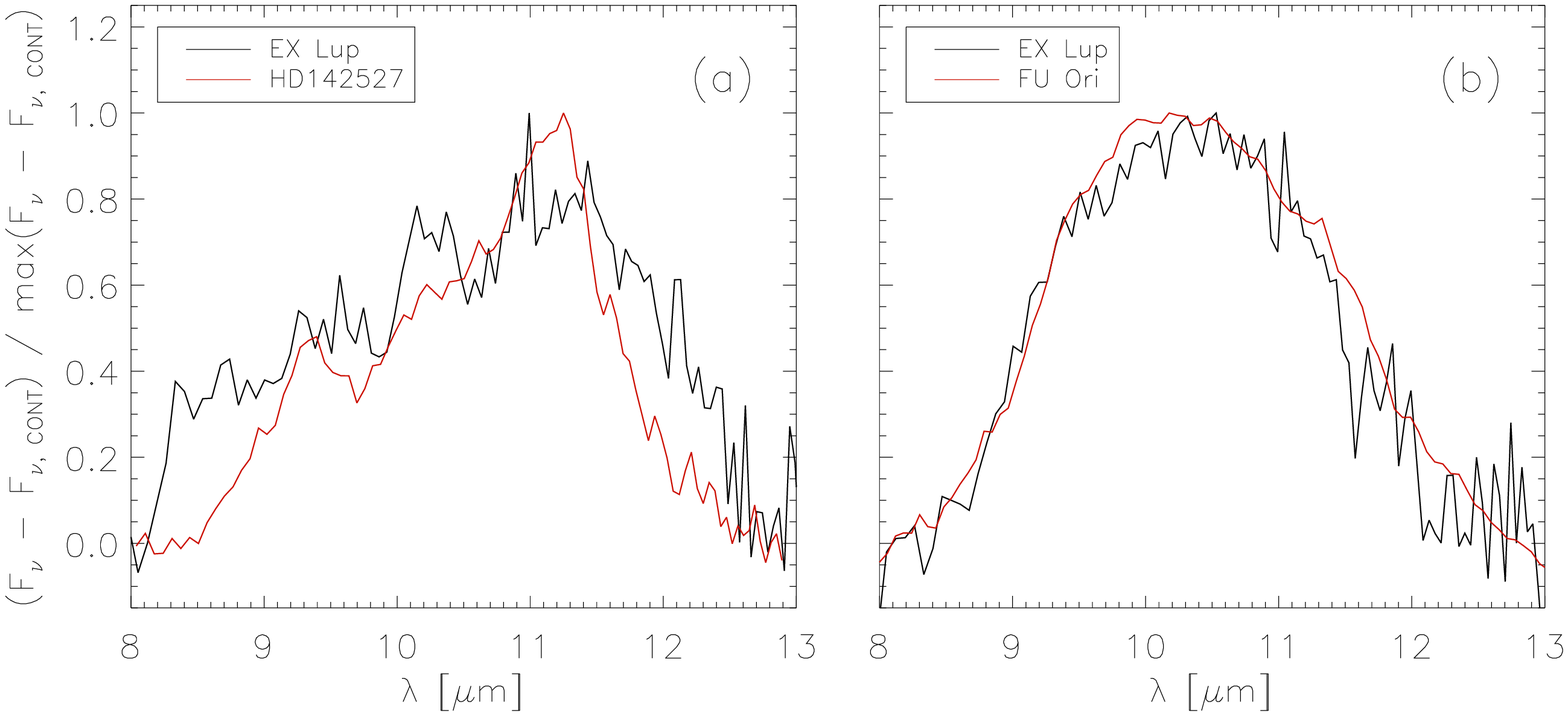}
\caption{VLTI/MIDI spectra of EX\,Lup using the U1U4 configuration with 121\,m projected baseline length. {\it Left:} Correlated
spectrum (inner disk) of EX\,Lup (black solid line) resembles that of the inner disk spectrum of HD142527 (red solid line, 
\citet{ref:van_boekel2004}) indicating very  high crystallinity. {\it Right:} The outer disk spectrum of EX\,Lup is very similar
to the Spitzer IRS spectrum of FU\,Ori, for which it was shown that large grains (several {\micron} in size) are needed to 
explain the shape of the 10\,{\micron} feature complex.}
\label{fig:comp_spec_inner_disk}
\end{center}
\end{figure}

In the Spitzer IRS spectra, the disk of EX\,Lup remains unresolved allowing us to investigate the integrated spectrum
of the disk as a whole only. To obtain information about the spatial distribution of the newly formed crystals we used MIDI
on the VLTI. In our first measurement in the U2U3 configuration (22 Jun 2008) with a projected baseline length of 37.4\,m EX\,Lup remained
spatially unresolved. The second measurement in the U1U4 configuration on 16 July 2008 (projected baseline length 121.2\,m), however, 
resolved the source. In the following we will use the naming convention of \citet{ref:van_boekel2004} and we will
refer to the correlated spectrum as "inner disk spectrum". We also subtracted the correlated spectrum from the total spectrum and
the resulting uncorrelated emission will be called the "outer disk spectrum". 
In Fig.\,\ref{fig:comp_spec_inner_disk}\,{\it Left} we present the inner disk spectrum of EX\,Lup. The spectrum peaks at 11.3\,{\micron} 
which corresponds to the wavelength of the strongest forsterite peak in the 8--13\,{\micron} region. The inner disk spectrum of EX\,Lup
resembles that of HD\,142527 which is known to originate in regions with $\gtrsim90\,\%$ crystallinity \citep{ref:van_boekel2004}, suggesting also very high 
crystallinity in the EX\,Lup inner disk spectrum. The outer disk spectrum of EX\,Lup is shown in 
Fig.\,\ref{fig:comp_spec_inner_disk}\,{\it Right}. It is broader compared to the pre-outburst total spectrum and peaks at about 
10.2\,{\micron}. The feature is smooth and lacks any substructure, which would be the signature of crystalline silicates. 
Interestingly, the outer disk spectrum of EX\,Lup is very similar to the 10\,{\micron} feature complex of FU\,Ori \citep{ref:quanz2007}.
\citet{ref:quanz2007} have shown that the 10\,{\micron} feature of FU\,Ori is known to emerge from a region where significant grain growth 
has taken place, indicating also the presence of large (few {\micron}) grains in the outer disk atmosphere of EX\,Lup.

\begin{figure}[!ht]
\begin{center}
\includegraphics[scale=0.4, angle=0]{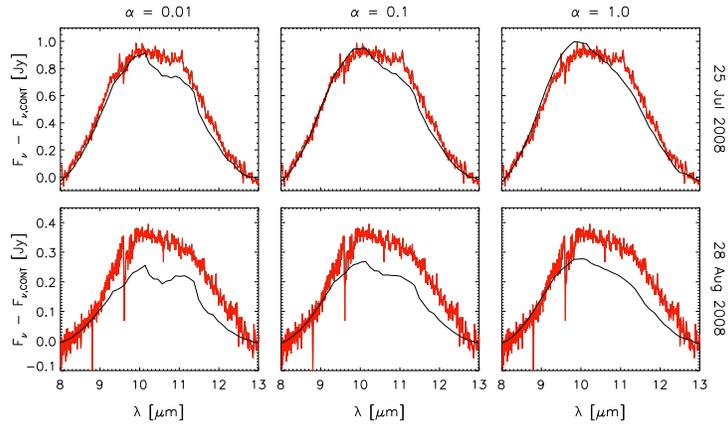}
\caption{Comparison of the VISIR observations (red solid lines) to our models A-AD (black solid lines). Our models reproduce the 
strength of the silicate feature very well on 25 Jul 2008, while on 29 Aug 2008 the model underestimates the flux. The shape of the
feature is reproduced the best on by models with $\alpha=0.1$, however, in Aug 2008 the shape of the spectrum is somewhat uncertain
around 9.8\,{\micron} due to the atmospheric ozone band. 
}
\label{fig:visir_fig}
\end{center}
\end{figure}

Two spatially unresolved spectra were taken with the VISIR instrument on VLT on 25 Jul 2008 and on 28 Aug 2008 (see Fig.\,\ref{fig:visir_fig}).
Although the VISIR spectra have better signal-to-noise ratio than those taken by the MIDI instrument, the quality of the spectra
is not as good as that of Spitzer IRS, allowing only moderate analysis of the spectral features. In July 2008, at the epoch of the 
first VISIR measurement the source was at about the same brightness level as in Apr 2008, when the outburst Spitzer IRS spectra was 
taken. The absolute flux level of the two spectra agrees well with each other, but the peak-to-continuum ratio is lower in the VISIR 
spectrum than than in the Spitzer IRS (1.6 vs. 1.4). The 10\,{\micron} feature complex seems to be broader in the VISIR spectrum, too. 
The broader and shallower features are usually interpreted as indication for the presence of larger, micron-sized grains. Similar
to the MIDI data the VISIR spectrum points towards the presence of large grains, too. The Aug 2008 VISIR spectrum
was taken at about the same optical brightness level as the first Spitzer IRS post-spectrum. Due to the lower signal-to-noise ratio
of the VISIR spectrum no significant difference can be found in the shape or the strength of the Spitzer and VISIR spectra above the noise level.


\section{Discussion}
\label{sec:discussion}


\subsection{Crystal formation}
\label{sec:crystal_formation}

\citet{ref:abraham2009} reported the in situ formation of forsterite crystals via annealing in the disk atmosphere of 
EX\,Lup during its 2008 outburst. The crystalline features we found in the 20--35\,{\micron} wavelength interval are 
also associated with crystalline forsterite, supporting the conclusion of \citet{ref:abraham2009}. 
It is interesting that no features of other crystalline dust species, apart from forsterite, can be seen in the mid-infrared 
spectra. For instance, enstatite (the magnesium-end member of the pyroxene solid solution series), which is frequently 
observed in protoplanetary disks (see e.g., \citealt{ref:van_boekel2005}, \citealt{ref:watson2009}, \citealt{ref:juhasz2010}), 
is not present in the spectra of EX\,Lup. Laboratory annealing experiments of amorphous 
silicates show that forsterite will form independently of the starting stoichiometry if the grains are small ($<1\,{\micron}$) 
and porous \citep{ref:fabian2000}. For larger (few {\micron}) and compact particles, annealing of amorphous grains with pyroxene
stoichiometry results in the formation of enstatite. Using the activation energies from \citep{ref:fabian2000} the annealing time of enstatite 
in the 1000-1300\,K temperature interval is about an order of magnitude longer than that of forsterite, but it decreases
rapidly with increasing temperature. At 1200\,K the calculated annealing time of enstatite is on the order of a minute. It is unlikely
that dust grains in the outburst are heated up to 1200\,K then cooled down below 1000\,K on a one-minute time-scale to prevent
the formation of enstatite.  
The only possibility for the formation of forsterite without enstatite production is that the parent amorphous grains
were sub-micron sized porous grains. Indeed, the analysis of the Spitzer IRS spectrum of EX\,Lup from the quiescent phase
show the presence of amorhphous grains with a mass-averaged grain size of 0.5\,{\micron} \citep{ref:sipos2009}. 

All crystalline silicate features, so far identified in protoplanetary disks, were associated with iron-poor, magnesium 
rich crystals (see e.g. \citealt{ref:van_boekel2005, ref:bouwman2008, ref:sargent2009} or \citealt{ref:henning2010} 
for a review). An important consequence of the absence of iron is the dramatic decrease of the optical and near-infrared
opacity of the grains. Iron-poor silicates are, therefore, cooler than ferromagnesian silicates at the same distance from the central star. 
Comparing the positions of the observed forterite features in the spectrum of EX\,Lup  with those of synthetic and natural olivines from 
\citet{ref:koike2003}, we estimated the iron content of the lattice to be less than 10\,\% with respect to magnesium. Such iron-poor crystals 
can be formed from iron-poor amorphous silicates, but also from ferromagnesian silicates if the annealing occurred under 
reducing conditions. 
Laboratory experiments showed that in the latter case the reduced iron will remain in thermal contact with the resulting crystals in 
the form of metallic inclusions \citep{ref:davoisne2006}. Such iron inclusions do not affect the positions of the forsterite bands while 
the grain as a whole will have higher optical to near-infrared opacity than that of isolated, pure forsterite crystals (see e.g. \citealt{ref:ossenkopf1992}).

Besides iron, amorphous carbon is the other important opacity source in protoplanetary disks. Carbon, as well as iron, has a featureless 
opacity curve and high optical to near-infrared opacity \citep{ref:jaeger1998}. \citet{ref:sipos2009} used amorphous 
carbon (20\,\% in terms of the total dust mass) in their radiative transfer modeling of the quiescent phase SED of EX\,Lup. Amorphous 
carbon, however, is not stable above 1000\,K as it oxidizes quickly and forms CO \citep{ref:gail2001}. Since annealing is expected to 
occur above 1000\,K, we do not expect amorphous carbon to be in thermal contact with silicate crystals. 

Based on the above constraints, we propose two scenarios to be tested for the formation of forsterite crystals in the disk of
EX\,Lup. In scenario A the parent grains were porous ferromagnesian silicates in thermal contact with carbon. As the outburst
set in and the temperature rose above 1000\,K, the silicate grains crystallized. The formation of CO from amorphous carbon
lowered the partial pressure of oxygen providing reducing conditions under which iron-poor crystals can form from ferromagnesian
silicates. To explain the observed variations of the strength of forsterite features an additional mechanism, e.g. 
radial/vertical mixing is required. 

In scenario B the parent amorphous silicate grains were iron-poor sub-micron sized porous grains in thermal contact with
carbon. As the temperature rose above 1000\,K at the beginning of the outburst, the silicate grains were annealed to crystalline 
forsterite, while the amorphous carbon oxidized and formed CO. Due to the low opacity of the crystals (in the absence of carbon)
their temperature decreased. This resulted in the weakening of the forsterite features in the 10\,{\micron} region while in the 
same time the features in the 20--35\,{\micron} region became stronger, which might explain the observed spectral variations. 


\subsection{Modelling}

We modeled the SED, the mid-infrared spectra and the interferometric visibilities simultaneously using the 2D radiative transfer (RT) code 
RADMC \citep{ref:dullemond_dominik2004a} and a 1+1D turbulent mixing code \citep{ref:dullemond_dominik2004b}. 
The goal of our modeling was to find answers to the following three questions. Are the freshly produced crystals 
isolated iron-free particles or are they in thermal contact with iron?
Which physical process was responsible for the observed variation of the mid-infrared solid-state features? 
What kind of constraints can we put on the mechanism of the outburst? 
We calculated, therefore, seven series of models to simulate the two crystal formation scenarios (series A-, and B-), 
with (-AD-) and without (-NAD-) an optically thick accretion disk and with different strength of the 
vertical turbulent mixing (-$\alpha$0.01, -$\alpha$0.1, -$\alpha$1.0). Details of the modeling procedure 
are described in Appendix A. We now briefly describe the modeling strategy in the different model series.

\begin{itemize}

\item[] {\bf Series A-AD-$\alpha$1.0, A-AD-$\alpha$0.1, A-AD-$\alpha$0.01.}
First, we started with a disk containing only amorphous silicates and amorphous carbon. The parameters of the starting disk 
model were taken from \citet{ref:sipos2009} and are summarized in Tab.\,\ref{tab:disk_params}. The inner radius of the
disk was taken to be 0.3\,AU (instead of 0.2\,AU as in \citealt{ref:sipos2009}) to get a better agreement between the model and the 
quiescent phase Spitzer IRS spectrum in the 5--10\,{\micron} wavelength interval. After the accretion rate has 
been set, we ran the RT code to obtain the temperature structure at the beginning of the outburst. Amorphous dust grains were 
replaced by forsterite grains with iron inclusions wherever the temperature in the disk exceeded 1000\,K. Then the turbulent mixing code 
was run to simulate vertical mixing over a 10-day period. Turbulent mixing was treated as a diffusion process using the $\alpha$ 
prescription (\citealt{ref:shakura_sunyaev1973}, see also Appendix A) for the diffusion coefficient. 
The strength of the turbulence was controlled via the value of $\alpha$, for which we used 0.01, 0.1 and 1.0. The last step 
was to update the density structure according to the results of the mixing code. The whole modeling cycle was then repeated. 
Accretion heating was simulated with an optically thick accretion disk as well as a hot spot on the stellar surface. 

\item[] {\bf Series A-NAD-$\alpha$1.0, A-NAD-$\alpha$0.1, A-NAD-$\alpha$0.01.}
Same as the A-AD- models, but no accretion disk was present in the models. The total accretion luminosity was assumed to
be radiated away from the hot spot. 

\item[] {\bf Series B-AD.}
We started the modeling in the same way as in the A-AD- models. Amorphous dust grains were replaced with isolated iron-free 
forsterite crystals wherever the temperature in the disk exceeded 1000\,K. Amorphous carbon was also removed from the dust 
phase above 1000\,K. In this model series we did not use the vertical mixing code, we only updated the accretion rate
and calculated models on each 10th day. 
\end{itemize}

For the amorphous grains we applied the dust model of \citet{ref:sipos2009}, which was used to model the quiescent phase SED. During the 
modeling we used two grain models for forsterite crystals. For isolated iron-free forsterite crystals we used mass absorption 
coefficients calculated from the optical constants using Distribution of Hollow Spheres (DHS, \citealt{ref:min2003}) scattering theory.
We used the optical constants of \citet{ref:sogawa2006} longwards of 2\,{\micron} and supplemented them with that of 
\citet{ref:scott_duley1996} shortwards of 2\,{\micron}. For composite grains, where forsterite and iron are in thermal contact 
with each other, we calculated the optical constants of the aggregate from those of the constituents first, using effective medium 
theory (APMR mixing rule, \citealt{ref:min2008}). For the constituents of the aggregate we used the DHS polarizability. Finally 
Mie theory was applied to calculate the mass absorption coefficient of the aggregate. The same optical constants were used for forsterite 
and for the isolated crystals, while for iron we used the optical constants of \citet{ref:ordal1988} between 0.66\,{\micron} 
and 285\,{\micron} and extended it with the optical constants of \citet{ref:pollack1994} shortwards of 0.66\,{\micron} and 
longwards of 285\,{\micron}. 

\begin{figure}[!ht]
\begin{center}
\includegraphics[scale=0.6]{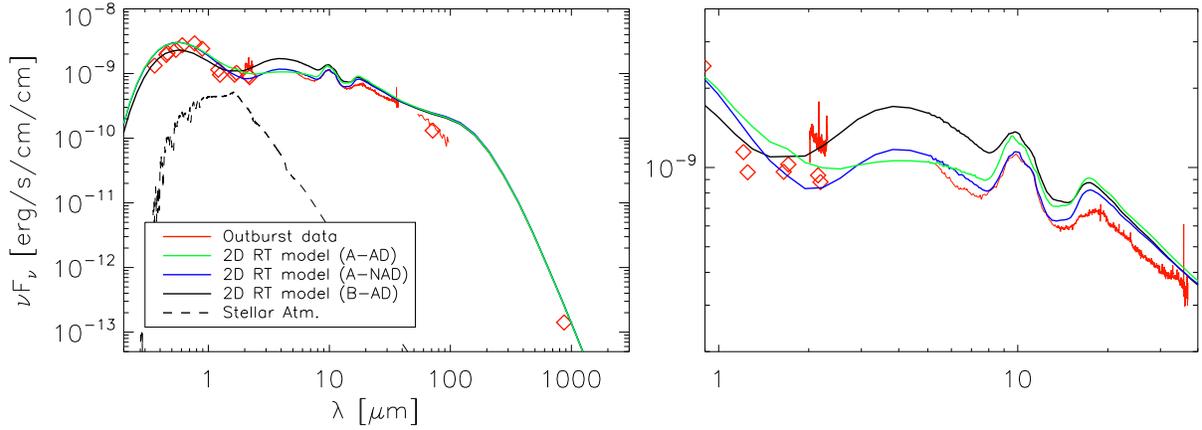}
\caption{{\it Left:} SED of EX\,Lup in the 2008 outburst. Data shown with red are measured within 24 hours (on 20 and 21 Apr 2008).
The black dashed line shows the M0 stellar emission. {\it Right:} In the 1.5--20\,{\micron} wavelength interval the A-NAD
models fit the observed data the best, while outside of this wavelength range the quality of the fit is very
similar in all three model families.}
\label{fig:model_SED}
\end{center}
\end{figure}

In Fig.\,\ref{fig:model_SED} we compared the observed SED of EX\,Lup in the outburst to model calculations. 
Although our models reproduced the general shape of the observed SED, there were quantitative differences. 
The largest differences between our models and the observations occured in the 1--8\,{\micron} wavelength interval (A-AD models) 
and in longwards of 60\,{\micron}. The former is related to the presence or absence of an optically thick accretion disk 
inwards of 0.3\,AU. We will discuss this problem in Sec.\,\ref{sec:outburst_mechanism}. 

The most likely explanation why our models overestimated the observed fluxes longwards of 60\,{\micron} is that the disk of EX\,Lup 
was not in an equilibrium state.  RADMC, as well as other RT codes used to model protoplanetary disks, calculates the temperature structure 
of the disk in radiative equilibrium. The emission longwards of 60\,{\micron} is dominated by the disk interior which is optically thick 
to the heating radiation of the star and the disk atmosphere above it. The high optical thickness prevents the disk interior to react
to the variable irradiation luminosity quickly. Since the optical depth in the far-infrared to millimeter wavelengths is lower than at shorter 
wavelengths the observer can see deeper into the disk observing those regions, that are still not accommodated to the changed irradiation luminosity.  
The optically thin, upper layers of the disk can, however, react to the variable irradiation luminosity almost immediately.


\subsubsection{Radial and vertical mixing of solids}
\label{sec:mixing}

\begin{figure}[!ht]
\begin{center}
\includegraphics[scale=0.4, angle=0]{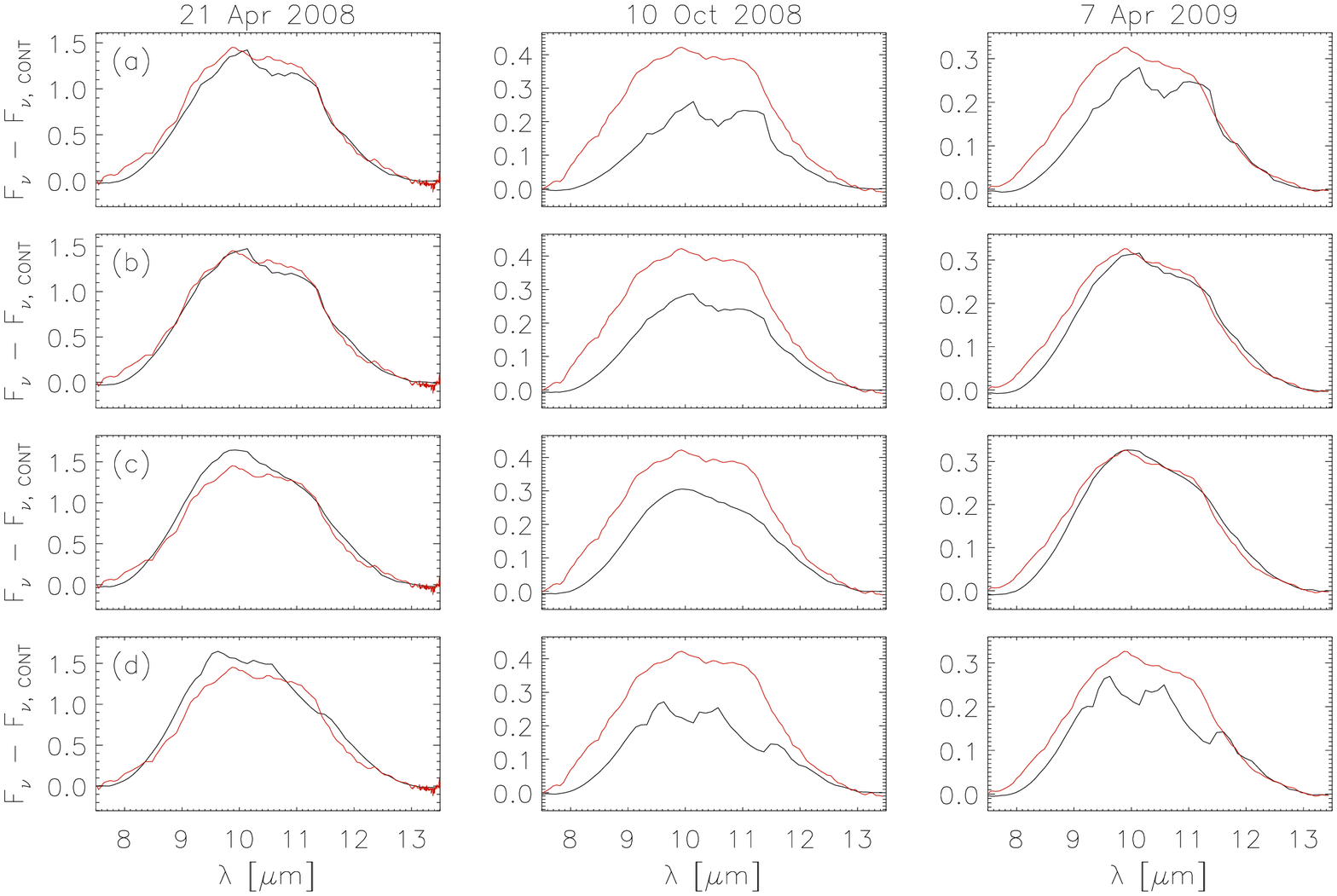}
\caption{Comparison of the observed 10\,micron silicate features with our model predictions. 
The observed Spitzer IRS spectra shown with red, while the model predictions are drawn
with black. Each column in the figure shows the observations and result of the modeling at an epoch
shown at the top of the columns. The different rows indicate different model series; ({\it a}) Model A-AD-$\alpha0.01$, 
({\it b}) Model A-AD-$\alpha0.1$, ({\it c}) Model A-AD-$\alpha1.0$, ({\it d}) Model B-AD. }
\label{fig:model_spec_short}
\end{center}
\end{figure}

The first question we tried to answer with our modeling was which of the two crystal formation scenarios, outlined in
Sec.\,\ref{sec:crystal_formation}, was at work in the disk of EX\,Lup. In Fig.\,\ref{fig:model_spec_short} we compared the
calculated mid-infrared spectra of models A-AD\footnote{The spectra of models A-AD and A-NAD are very similar to each
other, therefore, to avoid confusion we show only the spectra of models A-AD.} and B-AD. As can be seen in 
Fig.\,\ref{fig:model_spec_short}a-c, the positions of forsterite bands in the observed and modelled
spectra matched very well in models A-AD. The strengths of the forsterite peaks are dependent, however, on the strength of the
vertical mixing. The positions of the forsterite peaks in the B-AD models were shifted systematically towards shorter wavelengths 
compared to the observations (see Fig.\,\ref{fig:model_spec_short}d). The shift of the peak position was certainly related
to temperature/radiative transfer effects since in the mass absorption coefficients there was no difference in the peak
positions of isolated forsterite crystals and forsterite-iron aggregates. Moreover spectra of B-AD models showed also stronger
forsterite features, especially in the post outburst epochs, compared to the observed spectra. Thus, our modeling suggests
that the freshly produced forsterite crystals had to be in thermal contact with iron, which excludes scenario B as a crystal
formation mechanism. 

The next question we addressed was whether vertical mixing can explain the observed variation in the shape
of the mid-infrared solid-state emission features. Without considering vertical mixing, the apparent crystallinity in spatially 
unresolved spectra depends on the ratio between the size of the crystallized region and the total (crystalline + amorphous) 
silicate emission zone seen at 10\,{\micron}. The former depends on the irradiation luminosity at the peak of the outburst 
and does not change with time. The total silicate emission zone, however, scales linearly with the actual irradiation 
luminosity (see, e.g., \citealt{ref:kessler-silacci2007}). If the irradiation decreases, the total silicate emission would shrink
accordingly. 
Since the size of the crystalline region remains unchanged, the apparent crystallinity increases. 
So far we assumed that in the crystalline region the value of crystallinity is 1.0, i.e., every dust grain is in crystalline
form and this does not change with time. The effect of vertical mixing is to decrease the crystallinity with time in the 
crystalline zone by stirring up amorphous grains from deeper layers of the disk and, at the same time, mixing down crystals. 
In this way vertical mixing might compensate the effect of the shrinking total silicate emission zone. 

\begin{figure}[!ht]
\begin{center}
\includegraphics[scale=0.4, angle=0]{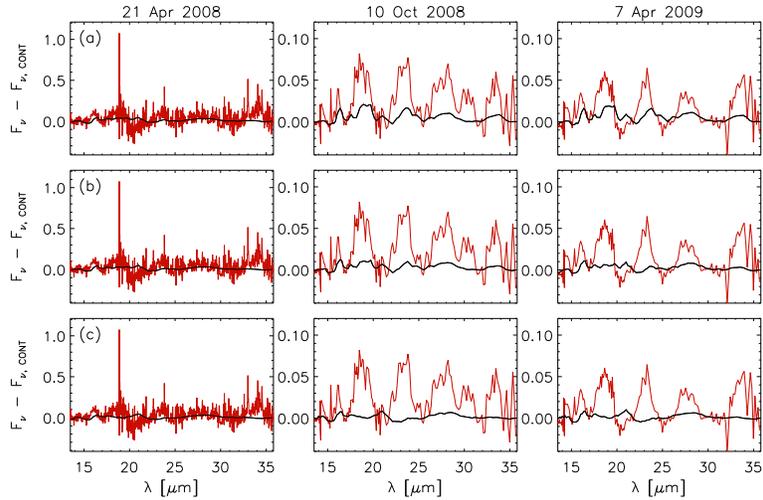}
\caption{Same as  Fig.\,\ref{fig:model_spec_short}, but for the 13--35\,{\micron} wavelength interval.}
\label{fig:model_spec_long}
\end{center}
\end{figure}

In Fig.\,\ref{fig:model_spec_short}a-c we compared the observed spectra in the 7.5--13.5\,{\micron} region 
with those of models A-AD with different values of $\alpha$. In our weakest mixing case ($\alpha=0.01$, Fig.\,\ref{fig:model_spec_short}a), 
our models predicted stronger forsterite features than observed. In contrast, models with the strongest possible mixing
($\alpha=1.0$, Fig.\,\ref{fig:model_spec_short}a), predicted too weak crystalline features compared to our observations. 
The best agreement between our models and the observed spectral features in the 7.5--13.5\,{\micron} was achieved using $\alpha=0.1$. 
The strength of the 10\,{\micron} feature complex is well reproduced for the outburst (Apr 2008) and the second post-outburst 
spectra (Apr 2009), but our model overestimates the feature strength compared to the spectrum taken in Oct 2008. 
The disagreement between the model and the observed spectra is about 35\%. Since the emission at the 10\,{\micron} is optically 
thin it reacts to the changing irradiation immediately and the infrared radiation is proportional to the irradiation flux. This means that the 
35\% difference between our models and the observed flux level translates to a difference of $\sim$0.3 mag difference in the optical. 
Since we had no optical photometry at the time when the Spitzer IRS spectrum was taken, only a few days after, the difference could be caused 
by the extrapolation to the optical brightness used in our modeling. 
In Fig\,\ref{fig:model_spec_long}a-c we compared our models to the observed spectra in the 13.5--36\,{\micron} wavelength
interval. Although our models with $\alpha=0.1$ matched the observed spectra in the 7.5-13.5\,{\micron} region well, 
the observed forsterite features longwards of 20\,{\micron} are far stronger than in any of our models independently of
the value of $\alpha$ (see Fig.\,\ref{fig:model_spec_long}).

Two explanations could be used to solve the discrepancy between the observed and modeled crystalline features longwards of 
20\,$\micron$. A solution could be that, in reality, the size of the crystallized region was higher than in our models. This can be due 
to a higher peak luminosity of EX\,Lup in the outburst or due to a lower crystallization temperature. In this case, 
the outburst spectrum could have contained some weak forsterite emission features longwards of 20\,{\micron}, which were hidden by
the noise in the measured spectra. The uncorrelated MIDI spectrum, however, excludes this possibility. We tested this scenario with our 2D RT model.
If the crystallized region is larger than about 1.5--1.8\,AU, the calculated uncorrelated spectra of our models show strong forsterite bands, 
which are not observed in the MIDI spectra.  Moreover, models which reproduced the strength of the forsterite features longwards
of 20\,{\micron} in the spatially unresolved post-outburst spectra, overestimated the strength of the crystalline features in the 10\,{\micron} region.
Using our modeling framework with vertical mixing, we were not able to reproduce the strength of the crystalline features both shortwards and longwards 
of 20\,{\micron} in the post-outburst spectra, and the spatially resolved mid-infrared interferometry simultaneously.

The other explanation would be that forsterite crystals were not present in the disk atmosphere outwards of 1.5--1.8\,AU 
in July 2008, but they were transported to a few AU by October 2008. The transport of crystals over a 2\,AU distance within three 
months implies an average radial velocity of about 38\,km/s. Radial mixing is far too slow for such a transport velocity. Instead, 
it is likely that in this scenario the crystals are transported by e.g. a wind driven transport mechanism. 
Indeed, optical and near-infrared spectroscopic observations suggest the presence of a wind during the outburst of EX\,Lup 
( \citealt{ref:aspin2010,ref:goto2011,ref:kospal2011}, Sicilia-Aguilar et al., in prep). The quantitative
modeling of this scenario is, however, beyond the scope of this paper.

\subsubsection{Outburst mechanism}
\label{sec:outburst_mechanism}

EXor outbursts are usually considered to be a 'down-scaled' version of FUor outbursts, suggesting the
same triggering mechanism of the outbursts in the two classes \citep{ref:hartmann_kenyon1996}.
The most widely accepted model of FUor outbursts is the self-regulated thermal instability (TI, \citealt{ref:bell_lin1994}).  
Although being the most widely accepted model, even TI cannot explain all observational properties of FUors (see e.g. \citealt{ref:zhu2007}).
In case of TI the released viscous energy in the disk midplane is trapped in the disk above a certain 
threshold surface density value. The radiative trapping is caused by the opacity of H$^{-}$ ions in the 
partially ionized disk regions. The effective viscosity as well as the mass transfer rate in the disk within
the ionized radius increases. The ionization front propagates outwards in the disk up to a certain radius then
retreats again. The gas inwards of the ionization front is accreted onto the central star. 

\begin{figure}[!ht]
\begin{center}
\includegraphics[scale=0.4, angle=0]{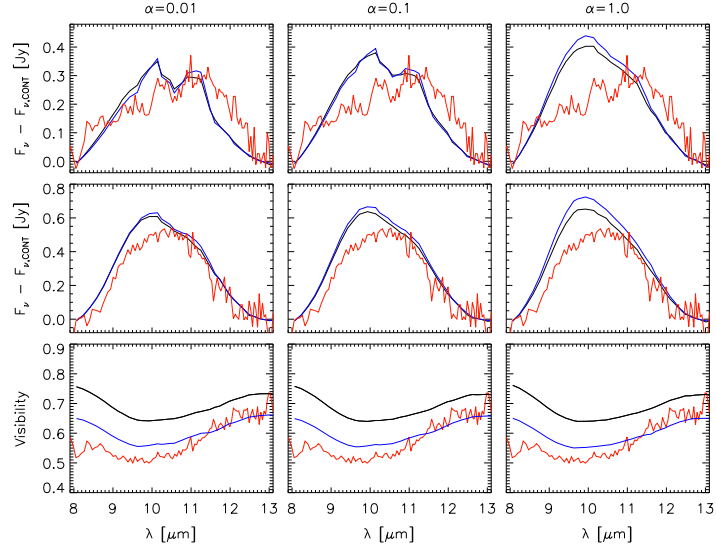}
\caption{Comparison of MIDI observations (red solid lines) to our models A-AD (black solid lines)
and A-NAD (blue solid lines). {\it Top:} Correlated (inner disk) spectrum, {\it Middle:} Uncorrelated (outer disk) spectrum, 
{\it Bottom:} Spectrally resolved visibilities. The different columns show models with different values of $\alpha$ used for the mixing 
simulation. 
The correlated spectra are reproduced the best with $\alpha=0.01$. The difference between our models and the observed
outer disk spectra can be explained with the presence of large (few {\micron}) grains in the disk atmosphere, which are not included in our models.
Although the shape of the silicate feature is only marginally sensitive to the presence of an optically thick accretion disk, 
the visibilities are more consistent with models without an accretion disk. }
\label{fig:vis_fig}
\end{center}
\end{figure}

In the following we will investigate the TI model in the context of our observations. 
TI requires an optically thick accretion disk (due to the presence of H$^{-}$ ions) 
which dominates the SED of FU Ori-type objects at optical and near infrared wavelengths (see e.g. \citealt{ref:zhu2007}). 
The comparison of our A-AD and A-NAD models to the observed SED of EX\,Lup showed that models without an optically thick accretion 
disk fit the SED better in the 1--8\,{\micron} wavelength range (Fig.\,\ref{fig:model_SED}).  In order to quantify the 
goodness of the SED fit we calculated and compared the $\chi^2$ of the fit for all model series.
The number of tuned parameters were only two, the value of $\alpha$ (except for B-AD models) and whether or not an optically
thick accretion disk was present within the inner hole of the dust disk.. The number of fitted data points were 1435. The $\chi^2$ values for the B-AD model was 3409. The 
A-AD-$\alpha$0.01, A-AD-$\alpha$0.1 and A-AD-$\alpha$1.0 models resulted  in a $\chi^2$ of 1423, 1372,1290, respectively. 
Finally the A-NAD-$\alpha$0.01, A-NAD-$\alpha$0.1 and A-NAD-$\alpha$1.0 models provided a $\chi^2$ of 533, 470 and 492, respectively.
It can be easily seen that the goodness of fit is very sensitive to the presence (or rather to the lack) of an optically thick accretion 
disk, while only marginal dependence is visible on the value of $\alpha$. Independently of the value of the viscosity 
parameter, the best fit model family is the A-NAD, while the absolute best fit model is A-NAD-$\alpha$0.1
Mid-infrared visibilities of models without an optically thick accretion disk gave also better agreement with the observations (see Fig.\,\ref{fig:vis_fig}).
This suggests that there was either no optically thick accretion disk present during the outburst of EX\,Lup, or the emission 
at optical wavelengths came from a narrow ring of hot gas in the inner edge of the gas disk. 


One of the main differences between FUor and EXor outbursts is the time-scale over which
the outburst occurs. The TI model {\it alone}, without any additional perturbation, is not able to reproduce
the rise-time in the case of FU\,Ori and of V1515 Cygni \citep{ref:clarke2005}. In these cases the brightening happened
over about a year, which is an order of magnitude longer than in the case of EX\,Lup (few weeks). 
The relatively short duration time of the outburst of EX\,Lup puts constraint also on the outer radius, from 
where gas can be accreted onto the star. The mass transfer in the disk occurs by viscous diffusion, 
the time-scale of which can be estimated using 
\begin{equation}
t_{\rm visc}\approx \frac{r^2}{D}\approx\frac{r^2}{H_p^2}\frac{1}{\alpha\Omega},
\label{eq:visc_tscale}
\end{equation}
where $D$ is the diffusion coefficient ($D=\alpha c_s H_p$), $r$ is the radius, $H_p$ is the pressure scale 
height of the gas and $\Omega$ is the Kepler frequency. We used  Eq.\,\ref{eq:visc_tscale} to estimate the 
outer radius from which material can be accreted within 10 months. We assumed that $r/H_p=10$ and for $\alpha$ we
used the highest theoretically possible value, $\alpha=1.0$. The resulting radius is $\approx 0.12$\,AU, which is well within 
the fitted inner radius of the dust disk (0.2--0.35\,AU, \citealt{ref:sipos2009}). This implies that the reservoir of 
material for the 2008 outburst should have been present \emph{in the inner hole of the dust disk}. 
This calculation implies that the hole in the disk, observed in the quiescent phase SED, should not be empty, but it is 
filled up with gas, which is optically thin in the continuum. On the other hand, if thermal instability takes place in the disk, 
the disk should be optically thick due to the presence of H$^{-}$ ions, which should be visible in the SED.

The mass of gas accreted onto the star during this outburst was about 10$^{-7}$\,M$_\odot$, and the calculation above implies 
that all of this gas should be located within 0.12\,AU from the central star. If we assume that such high amplitude outburst,
like that in the 1950s or in 2008, consume all the available gas within 0.12\,AU a mass transfer rate of $2\cdot 10^{-9}$\,M$_\odot$/yr 
or higher is required to maintain the observed amplitude and frequency (one in every 50 years) of these large eruptions.
The estimated mass transfer rate is a lower limit, as we neglected all the smaller amplitude outburst, which occurred since the 1950s, 
but it is more than an order of magnitude higher than what \citet{ref:sipos2009} estimated. The required mass transfer 
rate in the disk for TI to operate was found to be $5\cdot 10^{-7}$\,M$_\odot$/yr or higher \citep{ref:bell_lin1994}, which is far
higher than the estimated accretion rate in the quiescent phase of EX\,Lup.  If the outbursting region is fed with a lower
rate, thermal instability \emph{does not occur}, the accretion onto the star is stable and steady.
The equilibrium curves calculated by \citet{ref:bell_lin1994} also suggests that between $5\cdot 10^{-7}$\,M$_\odot$/yr and 
about $10^{-5}$\,M$_\odot$/yr there is no stable solution for the accretion. If thermal instability sets in, the disk
quickly flips to the outburst phase and the accretion rate increases to above $10^{-5}$M$_\odot$/yr. 

The mass transfer rate \emph{in the disk} is, however, not necessarily the same as the accretion rate \emph{onto the star}. 
The accretion rate onto the star can be lower than the mass transfer rate in the disk due to, e.g., interaction between the stellar 
magnetic field and the accreting gas (see, e.g., \citealt{ref:romanova2002, ref:romanova2004}, D'Angelo et al. in prep), which causes the 
gas to pile up in the disk. The gas lines, used as accretion indicator, are formed in the funnel flow between the inner radius of the 
gas disk and the surface of the star (e.g. \citealt{ref:muzerolle1998}), therefore they \emph{only} measure the accretion rate 
onto the star. Thus, the mass transfer rate in the disk can be significant even if the accretion rate derived from, e.g., the Br\,$\gamma$ line 
is zero. In FUor outbursts, all gas piled up over 10$^4$ years (estimated time between two subsequent FUor outbursts) is accreted
onto the star since the accretion onto the star is as high as $10^{-4}$M$_\odot$/yr. In contrast, the accretion rate of EX\,Lup in the
outburst was three orders of magnitude lower. If the mass transfer rate in the disk of EX\,Lup was $5\cdot 10^{-7}$\,M$_\odot$/yr
(required for the TI to operate), EXor outbursts cannot accrete all piled up mass from the disk onto the star. In this case gas 
accumulates continuously in the inner regions of the disk and at some points the system should go into a FUor-type outburst,
meaning that EXors are in fact FUors in the "quiescent" phase. Since EX\,Lup has a disk mass of 0.02\,M$_\odot$ and no evidence
for a massive envelope has been found so far, the disk of EX\,Lup would be entirely consumed by only a very few FUor eruptions (assuming
$\dot{M}$=10$^{-4}$M$_\odot$/yr and an outburst duration of 50\,yr), which makes this scenario unlikely.
Our study suggests that the 2008 outburst of EX\,Lup is caused by increased accretion onto the central star as indicated
by the increase of the Br\,$\gamma$ line luminosity in the outburst compared to the quiescent phase. On the basis of the
observational data we conclude, however, that it is unlikely that the self-regulated thermal instability is the responsible mechanism 
triggering the outburst. 

 The recent work of \citet{ref:dangelo2010} could provide an alternative explanation for the outburst mechanism of EX\,Lup. 
If the magnetic truncation radius of a strongly magnetized star is close to the co-rotation
radius the accretion onto the star will not be continuous anymore, but becomes episodic. After the end of an outburst, in the quiescent
phase, the inner radius of the disk is located just outside of the co-rotation radius. The accretion onto the star is low and material piles 
up at the inner edge of the disk. The inner radius of the disk moves inwards and when it crosses the co-rotation radius a burst of accretion 
occurs that empties the piled up reservoir of material and moves the inner edge of the disk outside of the co-rotation radius again. 
The interesting property of this model is the shape of the accretion rate vs. time curve (see e.g. Fig.\,4 middle panel in \citealt{ref:dangelo2010}). The outburst
starts with a steep rise of the accretion rate which is followed by a slow plateau-like fading with quasi-periodic oscillations
on top of the slow decline. This is very similar to what is observed in the optical light curve of EX\,Lup, which is thought to be closely related
to the accretion rate (i.e. hot spot on the stellar surface). Not only the shape of the light curve, but also the predicted time-scales by these simulations
are in qualitative agreement with the observations.
For the parameters which are the closest to that of EX\,Lup (see Fig.\,5 in \citealt{ref:dangelo2010}) the simulations of 
\citet{ref:dangelo2010} predict that the duration of the outbursts is less than the viscous time-scale at the co-rotation radius, while the
time between two subsequent outburst is about ten times the viscous time-scale. To compare the observations to the simulation of \citet{ref:dangelo2010} we 
estimated the viscous time-scale at the co-rotation radius for EX\,Lup.  \citet{ref:sipos2009} measured $v\cdot sin(i)$ of EX\,Lup to be 4.4$\pm$2\,km/s. 
Using a stellar mass and radius of 0.6\,M$_\odot$  1.6\,R$_\odot$ \citep{ref:gras_velazquez_ray2005}, respectively, and assuming 20$^\circ$ inclination and 
$\alpha=0.1$ the resulting viscous time-scale is about three years. According to the models of \citet{ref:dangelo2010} the duty cycle of the outbursts
would be about thirty years while a single outburst would last less than three years. Although these numbers are close to the observed values, 
the agreement could only be considered qualitative, due to the large unceratainty in the value of $\alpha$ and the unknown inclination angle.
Despite of these uncertainties, the remarkable similarity of the observed light curve of EX\,Lup and the accretion rate curve predicted by 
\citet{ref:dangelo2010} suggests that a further quantitative comparison between this model and observations of EXors should be done.


\section{Summary and conclusions}
\label{sec:conclusions}

Several studies tried to find special characteristics of EXors, apart from the optical outbursts, which would make them distinguishable
from 'normal' T\,Tauri stars (e.g., \citealt{ref:herbig2007}, \citealt{ref:sipos2009}), but none were found so far. Thus, it was concluded 
that probably most, if not all T\,Tauri stars go through an eruptive phase during their evolution \citep{ref:vorobyov_basu2006}.
Despite the fact that studies of these objects provide information about young low-mass stars in general, little is known
about this phase of stellar evolution. In this paper we studied the disk and dust properties around EX\,Lup in its most recent 
outburst, in order to learn about dust processing and put constraints on the outburst mechanism. 

Our conclusions can be summarized as follows:

\begin{itemize}

\item The accretion rate of EX\,Lup increased to about $2.2\cdot 10^{-7}$M$_\odot$/yr on 21 Apr 2008. This is about three orders
of magnitude higher than in the quiescent phase \citep{ref:sipos2009}, indicating that the outburst is caused by increased accretion
onto the star. 

\item Modeling of the SED and mid-infrared visibilities suggests that more than 90\% of the total accretion luminosity is radiated away 
from a hot component which dominates the SED at optical wavelengths. This component can either be a hot spot on the stellar surface
or a narrow ring of hot gas at the inner edge of the dust disk. Our observations exclude the presence of an extended (R$_{\rm out}\approx0.3$\,AU)
optically thick accretion disk during the outburst.

\item Calculation of the viscous time-scale in the disk shows that all material accreted in the 2008 outburst of EX\,Lup
should have been located within 0.1\,AU from the central star. This is well within the inner hole of the dust disk (inner
radius: 0.3\,AU), suggesting that the inner dust-free hole is filled with gas, optically thin in the continuum.

\item We compared  accretion rates and time-scales as well as the SED of EX\,Lup and FUors. We concluded that, although
the triggering mechanism of the outburst of EX\,Lup is certainly related to accretion, it is unlikely that it is the same thermal
instability that is thought to operate in FUor systems.

\item The post-outburst Spitzer IRS spectra of EX\,Lup revealed the presence of several new crystalline features 
(19.5\,{\micron}, 24\,{\micron}, 28\,{\micron} and tentatively at 33\,{\micron}), which are associated with crystalline forsterite.  
The detection of these  forsterite features supports the conclusion of \citet{ref:abraham2009} that crystals
have formed in the disk atmosphere of EX\,Lup during the outburst. 

\item On the basis of the position of the forsterite bands (especially the 16\,$\mu$m band) we conclude that the lattice of the new crystals is
very iron-poor. On the other hand, the crystals should be in thermal contact with iron, otherwise their temperature would be 
too low. This suggests that the parent amorphous grains had to be also in thermal contact with iron either in the silicate
network or as metallic inclusions. 

\item Apart from forsterite, no other crystalline silicate species were identified in the mid-infrared spectra of EX\,Lup. 
In order to prevent the formation of crystalline enstatite the amorphous grains should be small ($<$1\,{\micron}) and
porous. The analysis of the quiescent phase Spitzer IRS spectrum of EX\,Lup indeed showed that the disk atmosphere contains small 
amorphous grains \citep{ref:sipos2009}.

\item The MIDI correlated spectrum of EX\,Lup shows very high crystallinity, while in the uncorrelated spectrum no crystalline
features are seen. This indicates that crystals are highly concentrated towards the center, which is reproduced in our models 
as crystal formation occurs only within 1.1\,AU from the central star.

\item The shape of the uncorrelated (outer disk) spectrum shows signatures of large grains (several {\micron}) in the disk atmosphere, 
which were not present before. This can only be explained by active vertical mixing, which stirred up these grains from deeper layers. 
It also implies that the turbulence in the quiescent phase is probably weaker than in the outburst, to allow these grains to settle 
below the disk atmosphere.

\item Our model predicts far weaker forsterite bands longwards of 20\,{\micron} in the post-outburst 
spectra, than observed in the Spitzer spectra, indicating the presence of forsterite beyond 1.1\,AU (i.e. lower temperatures). 
Crystals could not form in situ outwards of 1.1\,AU in the disk atmosphere, otherwise their feature would be visible in the MIDI 
uncorrelated spectrum. We concluded, therefore, that they were transported outwards from within 1.1\,AU by some wind-driven 
transport mechanism.

\item Our vertical mixing model could not reproduce the decrease in the strength of the forsterite bands in the 10\,{\micron}
silicate complex between October 2008 and April 2009. An alternative explanation could be if a significant fraction of the crystal 
population in the innermost 1.1\,AU was transported to larger radii. 

\end{itemize}

\acknowledgments
We thank the anonymous referee for the careful review of the manuscript that helped to improve the paper.
We thank the ESO staff for executing the observations in service mode. 
The research of \'A.\,K. is supported by the Netherlands Organization for Scientific Research.
The research of Zs. R. is supported by the 'Lend\"ulet' Young Researcher Program of the Hungarian Academy of Sciences.


\appendix

\section{Modeling of the outburst}
\label{sec:ap_modeling}
In order to obtain the temperature structure of the disk, we used  2D radiative transfer code RADMC \citep{ref:dullemond_dominik2004a}
This code was extensively tested against other
radiative transfer codes at low \citep{ref:pascucci2004} and high optical depths \citep{ref:pinte2009, ref:min2009}. 
Originally, only passive heating of the dust grains by the central star was considered in RADMC. For this study we also
include viscous heating by an accretion disk and also the emission from the hot spot on the stellar surface. 
Viscous heating was implemented as a one grid cell layer in the disk midplane, which radiates as a blackbody at 
the local effective temperature, assuming it is optically thick in the vertical direction. 
The effective temperature of the accretion disk at a given radius  ($T_{\rm eff}(r)$) is calculated according to 
\citet{ref:shakura_sunyaev1973}:
\begin{equation}
T_{\rm eff}(r)^4 = \frac{3GM_\star \dot{M}}{8\pi R_\star\sigma}\left[1-\left(\frac{R_\star}{r}\right)^{0.5} \right]
\end{equation}
Here, $G$ is the gravitational constant, $\dot{M}$ is the mass accretion rate, while $M_\star$ and $R_\star$ are 
the mass and radius of the star, respectively. 

We also consider the radiation of the hot spot where the funnel flow of the accreting gas is thought to reach 
the stellar surface. Assuming that the spot radiates as a blackbody, the temperature of the spot is given by 
\begin{equation}
 T_{\rm s}^4 = T_\star^4 + \frac{GM_\star\dot{M}}{f4\pi\sigma R_\star^3}\left[1-\left(\frac{R_\star}{r_{\rm tr}}\right)\right].
\end{equation}
Here, $T_\star$ is the effective temperature of the star and $f$ is the fraction of the stellar surface area covered 
by the hot spot. The quantity $r_{\rm tr}$ is the truncation radius from which the accreting gas is thought to move with free-fall 
velocity towards the surface of the star. For $r_{\rm tr}$ we assumed $r_{\rm tr}=5R_\star$ according to 
\citet{ref:gullbring1998}. The surface area of the spot ($f=0.5$) was determined by fitting the outburst SED at optical 
wavelengths. This value is higher than usually found in T\,Tauri stars (0.01, \citet{ref:calvet1998}). It was, 
however, also noted that the covering fraction of the hot spot can be as high as $\geq0.2$ in strongly accreting 
systems \citep{ref:calvet1998}. To account for the change of the covering fraction with the accretion rate we 
use the following (ad hoc) relation for $\dot{M}\leq2.2\cdot 10^{-7}M_\odot/{\rm yr}$:
\begin{equation}
f = 0.5\cdot\left(\frac{\dot{M}}{2.2\cdot 10^{-7}M_\odot/{\rm yr}}\right)^{0.5}
\end{equation}
for $\dot{M}>2.2\cdot 10^{-7}M_\odot/{\rm yr}$ we fixed the value of $f$ and used $f=0.5$. 

After we obtained the temperature structure of the disk at any instant, the procedure in \citet{ref:dullemond_dominik2004b}
was followed to investigate the turbulent mixing of solids in the disk time-dependently. The mixing of the solid is treated as 
a turbulent diffusion process and we also include dust settling. The conservation equation of dust grains we solve is
\begin{equation}
\frac{\partial f(m,z)}{\partial t} - \frac{\partial}{\partial z} \left[\rho_{\rm gas}D(m,z)\frac{\partial}{\partial z}\left( \frac{f(m,z)}{\rho_{\rm gas}}\right) \right] + \frac{\partial}{\partial z} \left(f(m,z) v_{\rm sett}(m,z) \right) = 0
\end{equation}
Here, $f(m,z){\rm d}m {\rm d}z$ represents the number of dust grains per square centimeter of the disk, $D(m,z)$ is
the diffusion coefficient and $v_{\rm sett}(m,z)$ is the settling speed. 
For the diffusion coefficient of the gas we used the standard $\alpha$ prescription
\begin{equation}
D_0=\alpha c_s H_p,
\end{equation}
where $c_s$ is the local sound speed and $H_p$ is the pressure scale height of the gas. The diffusion
coefficient of the dust ($D$) depends on how strongly the grains are coupled to the gas (i.e. the size of the grains), 
which is described by the Schmidt-number ($Sc$) as $D=D_0/Sc$. 
The Schmidt number is unity if the dust particles are well coupled to the gas, while for larger grains sizes
the decoupling is described by larger Schmidt numbers. During the simulation we assumed that $Sc=1$. 
The dust particles, which contribute to the 10\,micron emission feature of 
EX\,Lup, are mostly sub-micron sized for which the strong coupling is a good approximation. 
During the calculation we assume that the disk is vertically isothermal and we
calculate the local sound speed from the temperature at the disk midplane. The sound speed scales with
the square root of the temperature, therefore the error we made with the assumption of vertically isothermal
disk is  in the order of $(T_{\rm atm}/T_{\rm mid})^{0.5}$. At 1\,AU this factor is less than 2, which is basically
negligible compared to the uncertainty in the value of $\alpha$.

\bibliographystyle{aa}
\bibliography{ms}



\begin{deluxetable}{llrrr}
\tablecaption{Log of observations of EX\,Lup in our campaign, supplemented with a pre-outburst Spitzer IRS spectrum from the Spitzer Data Archive. 
For photometric observations F$_\nu$ is the derived photometric flux density for a given filter/wavelength, while F$_{\nu, {\rm corr}}$ is the corrected 
value for the emission lines (optical photometries only).}
\tablewidth{0pt}
\tablecolumns{5}
\tablehead{\colhead{Instrument} & \colhead{Date} & \colhead{$\lambda$ [{\micron}]} & \colhead{F$_\nu$ [mJy]} & \colhead{F$_{\nu, {\rm corr}}$ [mJy]}}
\startdata
SPITZER IRS   & 18 Mar 2005 & 5.5--38.0  &   -   &   - \\
WFI           & 20 Apr 2008 & 0.36 (U)   &  159$\pm$1.46  &  159$\pm$1.46\\
GROND         & 20 Apr 2008 & 0.45 (g)   &  450$\pm$62  &  293$\pm$62\\
WFI           & 20 Apr 2008 & 0.46 (B)   &  486$\pm$5 &  319$\pm$5\\
WFI           & 20 Apr 2008 & 0.54 (V)   &  554$\pm$5  &  419$\pm$5\\
GROND         & 20 Apr 2008 & 0.61 (r)   &  675$\pm$93  &  557$\pm$93\\
GROND         & 20 Apr 2008 & 0.77 (i)   &  825$\pm$114  &  754$\pm$114\\
GROND         & 20 Apr 2008 & 0.89 (z)   &  1025$\pm$141 &  721$\pm$141\\
SOFI          & 20 Apr 2008 & 1.21       &  462$\pm$13  &  -\\
GROND         & 20 Apr 2008 & 1.25 (J)   &  400$\pm$9  &  -\\
GROND         & 20 Apr 2008 & 1.65 (H)   &  529$\pm$13  &  -\\
SOFI          & 20 Apr 2008 & 1.71       &  586$\pm$16  &  -\\
GROND         & 20 Apr 2008 & 2.15 (K)   &  670$\pm$31  &  -\\
SOFI          & 20 Apr 2008 & 2.19       &  643$\pm$18  &  -\\
SPITZER IRS   & 21 Apr 2008 & 5.5--38.0  &   -   &   - \\
MIPS          & 21 Apr 2008 & 71.42      &  3130$\pm$220 &  -\\
LABOCA        & 21 Apr 2008 & 870        &  41$\pm$10   &  -\\
VLTI MIDI     & 22 Jun 2008 & 8.0--13.0  &   -   &  -\\
VLTI MIDI     & 16 Jul 2008 & 8.0--13.0  &   -   &  -\\
VLT VISIR     & 24 Jul 2008 & 8.0--13.0  &   -   &  -\\
VLT VISIR     & 25 Jul 2008 & 8.0--13.0  &   -   &  -\\
VLT VISIR     & 27 Jul 2008 & 8.0--13.0  &   -   &  -\\
VLT VISIR     & 28 Aug 2008 & 8.0--13.0  &   -   &  -\\
VLT VISIR     & 29 Aug 2008 & 8.0--13.0  &   -   &  -\\
SPITZER IRS   & 10 Oct 2008 & 5.5--38.0  &   -   &  -\\
SPITZER IRS   &  7 Apr 2009 & 5.5--38.0  &   -   &  -\\
\tableline
\enddata
\label{tab:logobs}
\end{deluxetable}

\begin{deluxetable}{lr}
\tablecaption{Parameters of the quiescent disk model from  \citet{ref:sipos2009}. 
Parameters in italics were adopted from the literature and kept fixed, while the
rest of the parameters were derived from fitting the SED.}
\tablewidth{0pt}
\tablecolumns{2}
\tablehead{\colhead{Parameter}&\colhead{Fitted value}}
\startdata
{\it Stellar temperature} 				& 3800\,K\\
{\it Stellar radius}            			& 1.6\,R$_\odot$\\
{\it Stellar mass}             			& 0.6\,M$_\odot$\\
{\it Visual extinction}      				& 0\,mag\\
{\it Distance}                   			& 155pc\\
Inner radius (dust disk) 			& 0.3\,AU \\
Outer radius (dust disk) 			& 150\,AU\\
Scale height (at the outer radius) 	        & 18\,AU\\
Flare index                             	& 0.09\\
Power exponent (surface density) 	        & -1.0 \\
Total disk mass (gas + dust) 		        & 0.025\,M$_\odot$\\
Inclination angle				& 20\,$^\circ$\\
\tableline
\enddata
\label{tab:disk_params}
\end{deluxetable}

\clearpage

\end{document}